# Securing IoT Devices in Smart Cities: A Review of Proposed Solutions

Andrés F. Betancur-López[1]

[1]Graduate Student, Universidad Pontificia Bolivariana, Escuela de Ingenierías, Medellín, Colombia
Email: andres.betancurl@upb.edu.co

**Abstract**

Privacy and security in Smart Cities remain at constant risk due to the vulnerabilities introduced by Internet of Things (IoT) devices. The limited computational resources of these devices make them especially susceptible to attacks, while their widespread adoption increases the potential impact of security breaches. This article presents a review of security proposals aimed at protecting IoT devices in Smart City environments. The review was conducted by analyzing recent literature on device-level security, with particular emphasis on lightweight cryptography, physically unclonable functions (PUFs), and blockchain-based solutions. Findings highlight both the strengths and limitations of current approaches, as well as the need for more practical, scalable, and resource-efficient mechanisms to ensure user privacy and data protection in IoT ecosystems.

**Keywords:** Internet of Things, security, privacy, Smart Cities, IoT devices

## Contents

## 1. Introduction

Smart Cities are transforming urban environments through innovations in Information and Communication Technologies (ICT), offering improved services and enhanced quality of life to their citizens. A central element of these innovations is the Internet of Things (IoT), which enables the interconnection of everyday objects via electronic devices that monitor, collect, and transmit data. This information supports decision-making processes, allowing governments and organizations to identify patterns, preferences, and behaviors that can be used to optimize public services (Rose, 2015; Van Zoonen, 2016; Zhang, 2017).

Despite these benefits, the widespread deployment of IoT devices also creates new risks to security and privacy. Data about individuals' movements, habits, health, and preferences can be illegally accessed, manipulated, or exploited by malicious actors (Rose, 2015, Elmaghraby, 2014; Longo, 2015). Safeguarding this information is therefore essential across the entire system—from the devices that capture data, to the communication networks, and finally the applications that process and deliver services (Khan, 2014).

IoT devices such as sensors and actuators are particularly critical because they constitute the origin of data flows (Kozlov, 2012). However, their small size, low energy consumption requirements, and limited computational capacity restrict the use of complex security algorithms (McKay, 2017). Furthermore, the rapid growth of the IoT market has encouraged many manufacturers, often with little experience in cybersecurity, to release inexpensive devices that are difficult to update or patch, leaving them permanently vulnerable (Federal Trade Commission Staff, 2015).

Most of the proposed solutions in the literature focus on network- and application-level protection. Yet securing the source of information—the devices themselves—remains a major challenge. If these devices are compromised, the entire system's security measures become ineffective in protecting user privacy.

This article reviews existing proposals to strengthen the security of IoT devices in the context of Smart Cities. By focusing on device-level approaches, the study addresses a gap in current research that has largely emphasized higher system layers. The review identifies three major trends: lightweight cryptography, physically unclonable functions (PUFs), and blockchain-based solutions. The remainder of this article is structured as follows: Section 2 describes the materials and methods used for the review; Section 3 presents and discusses the findings; and Section 4 provides concluding remarks and future directions.

## 2. Materials and Methods

We carried out a literature review of research articles on the security of Internet of Things (IoT) devices. To guide this process, we defined search criteria based on the problem described in the Introduction. Using these criteria, we built an initial search equation and applied it in several well-known databases. We then refined the equation step by step through manual screening of the first results.

With the final set of articles, we read and classified each one according to the type of security proposals they presented for IoT devices.

### 2.1. Search criteria

The idea of the Internet of Things dates back to the late 1990s, but its massive growth has come mainly from recent advances in Information and Communication Technologies (ICTs). Because of this, we focused on publications from the last ten years, paying particular attention to those published between 2017 and 2022. In addition, we applied the following criteria:

- The article must include either a specific proposal to secure IoT devices or a broader security framework that covers devices.
- It may focus on a particular sector or application, as long as it includes a security proposal for IoT devices.
- Only peer-reviewed research articles were considered.
- The publication had to be openly accessible to allow full review.

## 2.2. Data bases and Search Equation

To start the search, we selected the keywords IoT, device, security, and privacy. Since most work in this field is published in English, the search was conducted in that language. As we reviewed the first results, we identified additional useful terms — such as smart city, lightweight cryptography, physical unclonable function (PUF), and blockchain — and combined them with the initial keywords to refine the search.

The final search equation was:

> *TITLE-ABS-KEY ((iot AND device AND security AND privacy AND smart city) OR (iot AND device AND security AND privacy AND lightweight cryptography) OR (iot AND device AND security AND privacy AND (physical unclonable function OR puf)) OR (iot AND device AND security AND privacy AND blockchain))*

We applied this equation in the digital libraries listed in Table 1, searching within the titles, abstracts, and keywords of the publications.

**Table 1.** Data bases used for the search

| Data base | Type | URL |
|---|---|---|
| ACM Digital Library | Digital library | https://dl.acm.org/ |
| SpringerLink | Digital library | https://link.springer.com/ |
| Science Direct | Digital library | https://www.sciencedirect.com/ |
| Scopus | Digital library | https://www.scopus.com/home.uri |

## 2.3. Manual Screening and Selection of Publications

Once the initial results were obtained, the search equation was refined until it produced more specific and relevant outcomes for the study. On this preliminary set, we applied filters for publication year, type of publication, and access type, in line with the defined criteria.

The initial search returned a total of 4,483 results. We conducted a first screening of article titles, abstracts, and keywords to narrow down the selection. As highlighted in the Introduction and in the search criteria, the presence of the term device was a key factor. Applying this filter reduced the set to 93 publications, which were then inspected in greater detail to identify those most relevant to the study. This inspection involved carefully reading the abstract and introduction, along with a targeted review of the rest of the article's content. In the end, we selected 31 articles for detailed analysis.

Figure 1 illustrates the process of searching and selecting publications for this review. It shows how, as filters and criteria were applied, the set of sources was progressively refined to include only the most relevant contributions.

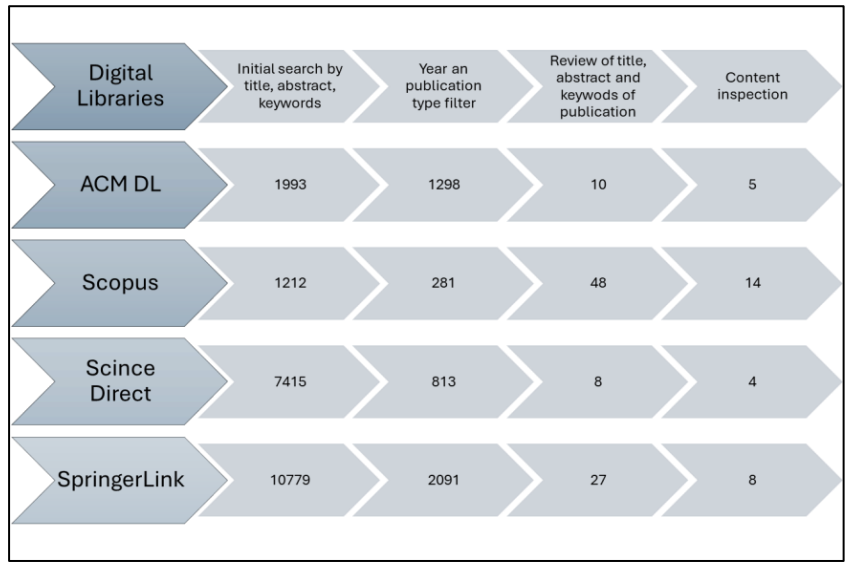

**Figure 1.** Publications search and selection process

## 3. Results and Discussion

Figure 2 illustrates the general architecture of IoT, which is typically organized into three layers: the perception layer, where sensors and actuators are located; the network layer, which provides internet connectivity for data transmission; and the application layer, where the collected data are used and from which actions are sent back to actuating devices (Witti, 2018). These devices face significant computational and energy constraints, which limit the feasibility of using complex or resource-intensive security schemes.

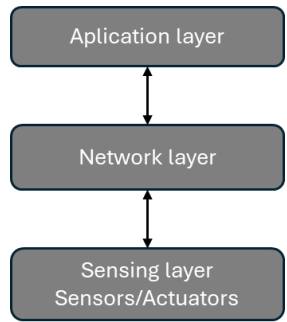

**Figure 2**. IoT architecture

In this section, we present the results of the reviewed articles, focusing on the proposals aimed at securing the perception layer. As shown in Figure 3, the analysis of the selected publications revealed three major trends in security approaches for IoT devices: lightweight cryptography, which appeared in 39% of the 31 studies reviewed; physical unclonable functions (PUFs), mentioned in 23% of the studies; and blockchain, reported in 19% of the studies. Other proposals included legal frameworks (6%) and alternative forms of authentication (13%).

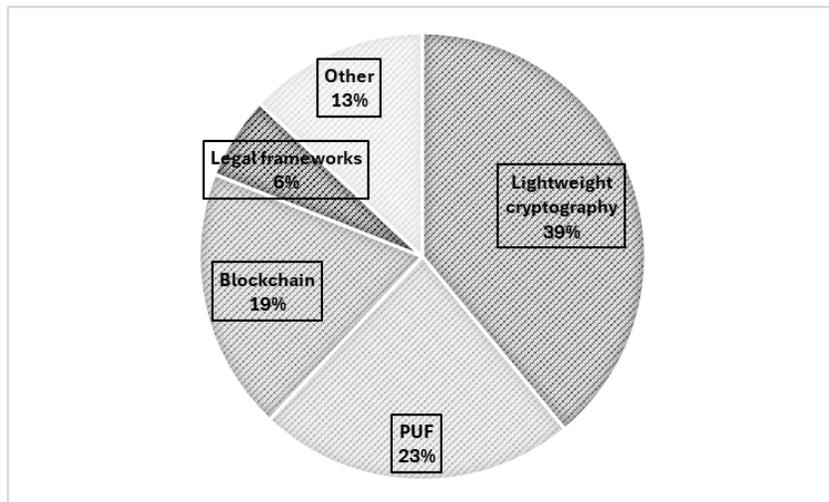

**Figure 3.** Trends in security approaches for IoT devices

## 3.1. Lightweight Cryptography

Cryptography involves the development and application of techniques to secure communication and data transfer between systems. It relies on complex mathematical calculations that typically require high computational resources. Considering the limitations of IoT devices, lightweight cryptography has been developed to secure them without exceeding their restricted capabilities. Figure 4 shows an example of an encryption algorithm.

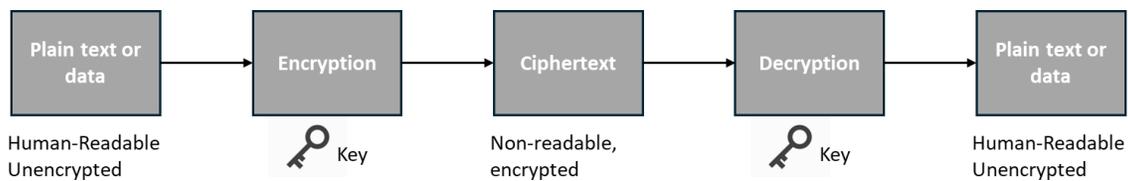

**Figure 4.** Cryptography

Among the lightweight cryptography proposals is the implementation of a compact block cipher. Block ciphers are algorithms that transform groups of bits (blocks) using arithmetic operations with a key. The algorithm proposed by Seo (2018) is based on addition, rotation, and XOR functions. It achieves compact implementation by adapting to word size, the number of general-purpose registers, and the instruction set of IoT devices. It was evaluated using the FELICS framework (Fair Evaluation of Lightweight Cryptographic Systems), achieving top performance rankings.

Another lightweight block cipher was proposed by Baskar (2015). Unlike the previous approach, it reorganizes addition, rotation, and XOR operations, and modifies key generation to enhance security. It builds on the Tiny Encryption Algorithm (TEA), which, as its name suggests, is a simple, lightweight algorithm well suited for resource-constrained devices.

Feistel ciphers are symmetric structures commonly used in block cipher design. Karakoç (2015) introduced a Feistel-based algorithm with alternating keys. This approach simplifies block cipher design by avoiding the need for a key schedule. Combined with the Feistel structure, it reduces memory usage and the number of operations, which translates into lower energy consumption. The proposed design strategy also decreases vulnerability to key-related attacks.

Similarly, Ragab (2020) and Li (2016) presented lightweight block cipher proposals, based respectively on XXTEA and on the Feistel structure. By modifying these algorithms, they sought to improve security while minimizing resource consumption, achieving measurable improvements compared to their original versions.

Biryukov (2017) introduced two categories of lightweight cryptography for IoT device security. The first, ultra-lightweight cryptography, prioritizes minimizing resource consumption to adapt to the severe restrictions of small devices. The second, ubiquitous cryptography, is more functional and intended for devices with greater—though still limited—computational capacity. This classification highlights the importance of tailoring algorithm design to the capabilities of the target devices.

Dhanda (2020) reviewed lightweight cryptographic algorithms, comparing block ciphers, stream ciphers, hash functions, and variants of elliptic curve cryptography (ECC). The comparison considered resource usage, efficiency, and energy consumption, among other factors. The review concluded that block ciphers and ECC are the most suitable solutions for IoT devices according to the evaluated criteria.

A modern cryptographic technique that represents an important advance is elliptic curve cryptography (ECC). It has gained wide acceptance because it is much harder to break than classical algorithms, due to the mathematical complexity of elliptic curves. Zhou (2019) proposed a lightweight ECC implementation on an 8-bit device, demonstrating that ECC can indeed be used in constrained environments. The smaller key sizes required by ECC translate into reduced memory and transmission requirements while maintaining high levels of security. Similarly, Rao (2020) proposed a lightweight ECC approach using two elliptic curves: one for key generation and another for encryption, thereby increasing security by separating the processes.

Mustafa (2018) reviewed various lightweight security techniques, focusing on lightweight cryptography and lightweight steganography. Steganography hides a message or data within another message or data, such as embedding information inside an image, video, or multimedia file. After analyzing different proposals, the study concluded that combining cryptography with steganography offers the best balance to achieve confidentiality, integrity, and authentication.

Considering the limited resources available, Schuß (2018) proposed the implementation of a cryptographic co-processor. This could be included in device design or added as a complementary module, in the form of a SIM card. Despite its small size, such a module can offer high storage capacity and offload computationally intensive security tasks from the main processor. It enables the use of traditional lightweight cryptographic techniques and allows straightforward hardware and software updates.

Finally, Zhang (2021) used cryptography to extract a digital fingerprint of IoT devices as a means of unique identification. A device fingerprint refers to a distinctive feature that allows its identification, which could be a serial number combined with the brand and model, or even its unique electrical response to a stimulus. Since these physical characteristics are never identical, the fingerprint is unique. The proposal involved applying cryptographic algorithms to extract and combine these features, thereby generating a digital fingerprint through the arithmetic transformations of the algorithm.

In summary, the reviewed studies show that lightweight cryptography has become a central strategy for securing IoT devices under strict resource constraints. Proposals range from compact block ciphers and Feistel-based structures to advanced techniques such as elliptic curve cryptography and combinations of cryptography with steganography. Other approaches include hardware-based solutions, such as cryptographic co-processors, and novel applications like device fingerprinting. Together, these contributions highlight the diversity of strategies being explored, while also underscoring the ongoing need to balance efficiency, energy consumption, and robustness against emerging threats.

## 3.2. Physical Unclonable Function (PUF)

Physically Unclonable Functions (PUFs) are techniques used to uniquely identify hardware elements, in this case IoT devices. During the manufacturing process of circuits and semiconductors, small physical variations occur in each component. When the circuits are exposed to a stimulus, their responses differ due to these variations. This unique response serves as a fingerprint that makes it possible to identify each device.

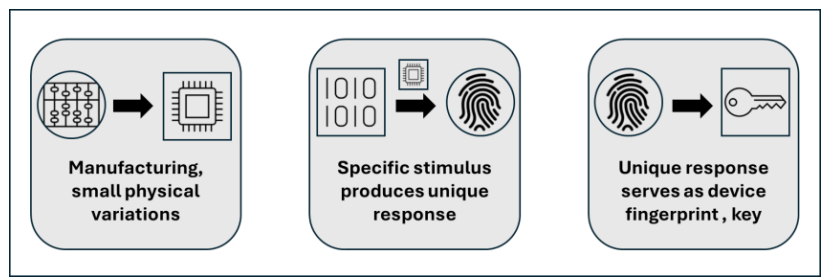

**Figure 5.** Example of Physically Unclonable Function (PUF).

Software-based security mechanisms become increasingly vulnerable as hardware capabilities advance and make it easier to break them. Hardware-based mechanisms provide an alternative, but they can still be compromised in cases such as theft or device cloning. PUFs address this challenge. Their compatibility with the limited resources of IoT devices gives them an advantage over other cryptographic solutions (Shamsoshoara, 2020). This study highlights these benefits, while also noting remaining challenges, particularly in extracting reliable device fingerprints from noisy electrical responses.

As noted above, PUFs enable the unambiguous identification of devices. Their main application in IoT security is device authentication, that is, verifying that a device's claimed identity within the network is genuine. Fenga (2018) proposed an authentication scheme based on PUFs, complemented with attestation — a mechanism to further verify the authenticity of a device. While most existing works focus only on authentication, this added layer of attestation strengthens IoT device security.

Huang (2020) also introduced an authentication mechanism. Its novelty lies in not relying on a single PUF to identify a device but instead creating an identity from the PUFs of all the integrated circuits (ICs) that compose the device. This approach makes authentication more robust and cloning more difficult, since changing even a single component alters the PUF response.

Another proposal combining PUFs with cryptography is that of Balan (2020). In this case, PUFs are used not only for device authentication but also for authenticating communication between circuits within the same device. Each circuit generates its own PUF, and the communication between them is encrypted. These PUFs can also be used to generate encryption keys for communication with other elements in the network.

Other studies have also explored PUF-based authentication from different angles. Satamraju (2020) proposed a protocol for mutual authentication between device and server using PUFs, with an additional third party serving as the authentication server. Sen (2020) suggested improvements in the challenges or stimuli presented to devices through PUFs, so that the responses are more efficient and improve authentication accuracy. Finally, Al-Aqrabi (2019) proposed the creation of an authentication entity for multiple actors in the IoT network. Given the diversity of interactions, the study also suggested using devices with dynamic partial reconfiguration capabilities to manage different connection sessions between devices.

In summary, Physically Unclonable Functions (PUFs) provide an effective hardware-based solution for uniquely identifying IoT devices and supporting secure authentication. Their strength lies in exploiting inherent manufacturing variations, which makes each device response unique and difficult to clone. The reviewed proposals highlight different applications, from basic authentication schemes to more advanced frameworks that combine PUFs with attestation, cryptography, or dynamic reconfiguration. However, challenges remain, particularly regarding noise in electrical responses and the practical deployment of PUFs at scale. Overall, PUFs represent a promising complement to lightweight cryptography, offering security mechanisms that are well aligned with the resource limitations of IoT devices.

### 3.3. Blockchain

In simple terms, blockchain is a system that maintains a distributed or shared database of assets or transactions, which is nearly impossible to tamper with. This characteristic is its main strength and the reason why blockchain can be applied in a wide variety of contexts. One such application is the security of devices in the Internet of Things (IoT).

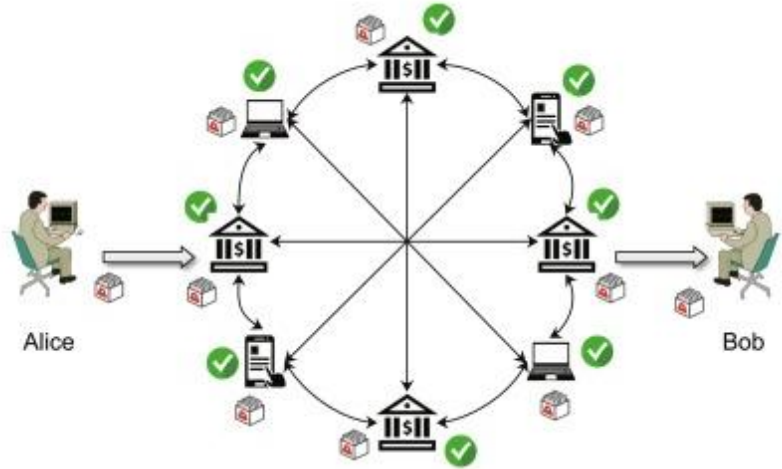

**Figure 6.** Blockchain (Aggarwal, 2021)

The most common proposal for securing IoT devices is authentication. As noted earlier, blockchain functions as a database, and each record can contain different pieces of information about a device. These data serve to identify the device and enable its authentication within the IoT network. Security is achieved because every transaction must be validated by a defined percentage of users in the network, each of whom maintains a copy of the original database. Once validated, the transaction is updated and distributed across all nodes.

In addition to these features, the stored data is encrypted and can only be accessed by the designated recipient. Furthermore, organizations may implement their own blockchain system using a permissioned model, in which a node must be authorized before joining the network, thereby adding another layer of security. Several studies, including those by Dehalwar (2022), Kolokotronis (2019), Kshetri (2017), and Gong (2021), present similar proposals in which blockchain is implemented for IoT device authentication, thus enhancing security by leveraging blockchain's inherent advantages.

Other studies take a different approach. For instance, Chanson (2019) observed that blockchain can be used to store IoT network data originating from or destined for the perception layer — that is, the data collected by sensors and the commands sent to actuators. However, this proposal does not address authentication. By contrast, Agrawal (2017) identified that blockchain can be used not only to store device identification data for authentication but also to secure the sensor and actuator data themselves. His study proposed using blockchain for both purposes, aiming to provide a more comprehensive security framework for IoT devices.

In summary, blockchain offers a decentralized and tamper-resistant framework that can significantly enhance IoT security. Most proposals focus on device authentication, ensuring that only legitimate devices can interact within the network. Other approaches extend its use to the protection of sensor and actuator data, broadening its role beyond identity management. While permissioned blockchain models further strengthen control and confidentiality, challenges remain in terms of scalability, energy consumption, and integration with resource-constrained IoT environments. Nevertheless, blockchain stands out as a promising complement to lightweight cryptography and PUFs, providing distributed trust and data integrity in IoT systems.

## 3.4. Other Proposals

Some of the reviewed publications present proposals that do not fall within the three major areas identified earlier. This is the case of Weber (2010; 2015), where the same author examined the legal framework for the Internet of Things at two different points in time, five years apart. His studies highlight persistent gaps in regulations or laws governing the new technologies associated with IoT, particularly in relation to the protection of personal data.

These works identify the need to establish a legal framework that spans the entire IoT chain — from device manufacturing to application deployment — in order to strengthen data protection. Such a framework would not only ensure accountability when system security or privacy is breached, but also act as a deterrent for attackers by imposing strict sanctions.

Other proposals focus on securing IoT networks as a whole, while still addressing device security. For example, Witti (2018) proposed a reference framework for system assurance based on policies and service-level agreements (SLAs) between applications. As long as the defined policies and SLAs are met, the system is considered secure, from sensors and actuators through to the applications that receive and process the data. The framework incorporates traditional techniques for device identification and authentication.

Mathur (2017) suggested an end-to-end solution in which any computer in the world could connect to an IoT device without the latter being directly exposed to the internet. This approach, similar to a VPN or tunneling system, adds an extra layer of security alongside traditional data authentication and encryption techniques.

Two additional proposals focus on device fingerprint monitoring as a mechanism to enhance security. Lauria (2017) proposed monitoring device network fingerprints, which include parameters such as IP address and related data (e.g., geolocation), MAC address, ports, operating systems, and others. The system continuously monitors these parameters and generates alerts or actions whenever changes are detected. Similarly, Elkanishy (2021) proposed monitoring Bluetooth connections. By parameterizing features such as distance between connected devices, number of devices, and signal strength, a unique fingerprint is created. Any changes in these monitored parameters trigger alarms and enable preventive measures to avoid security breaches.

In summary, although most efforts to secure IoT devices concentrate on cryptography, PUFs, and blockchain, a number of complementary approaches have also been explored. These include strengthening the legal and regulatory framework, defining service-level agreements as part of system assurance, creating secure end-to-end communication channels, and monitoring device fingerprints at the network or Bluetooth level. While these proposals are more diverse and sometimes sector-specific, they highlight the importance of addressing IoT security from multiple angles — legal, technical, and operational — in order to provide more comprehensive protection across the entire ecosystem.

## 4. Conclusions

The results of this review show that cryptography remains the predominant mechanism for securing IoT devices. As the most traditional technique, it has been adapted to the limitations inherent in IoT environments.

However, Physically Unclonable Functions (PUFs) have also been widely studied due to their strong compatibility with electronic devices. At the same time, there is growing interest in blockchain, a trend that is gaining momentum across many sectors thanks to its high levels of security and relatively straightforward implementation.

Cryptography will likely continue to be the most widely used mechanism, often in combination with other techniques. Nonetheless, there is a clear trend toward exploring new solutions, with blockchain emerging as a particularly promising option. As several studies demonstrate, combining lightweight techniques for authentication and encryption is both feasible and advantageous, as it enhances security by providing authentication, integrity, and confidentiality for IoT devices.

It is also important to note that while most efforts focus on protecting data and information against theft or manipulation, legal frameworks for security and privacy should not be overlooked. Such measures play a valuable deterrent role, reducing risks to IoT systems by holding actors accountable and imposing meaningful sanctions when breaches occur.

Overall, this review highlights that significant progress has been made, and ongoing research continues to explore alternatives for strengthening the protection of IoT devices. The ultimate goal is to safeguard not only the data they handle but also the individuals and organizations who depend on them.

## References

Aggarwal, S., & Kumar, N. (2021). Core components of blockchain. In S. Aggarwal, N. Kumar, & P. Raj (Eds.), *Advances in computers: Vol. 121. Introduction to blockchain* (pp. 193–209). Elsevier.

Agrawal, R., Faujdar, N., Chauhan, P., & Kumar, A. (2017). Security and privacy of blockchain-based single-bit cache memory architecture for IoT systems. *IEEE Access*.

Al-Aqrabi, H., Johnson, A. P., & Hill, R. (2019). Dynamic multiparty authentication using cryptographic hardware for the Internet of Things. Centre for Industrial Analytics, University of Huddersfield.

Balan, A., Balan, T., Cirstea, M., & Sandu, F. (2020). A PUF-based cryptographic security solution for IoT systems on chip. *EURASIP Journal on Wireless Communications and Networking*.

Baskar, C., Balasubramaniyan, C., & Manivannan, D. (2015). Establishment of lightweight cryptography for resource constraint environment using FPGA. In *Proceedings of the International Conference on Information Security & Privacy (ICISP 2015)* (pp. 11–12). Nagpur, India.

Biryukov, A., & Perrin, L. (2017). State of the art in lightweight symmetric cryptography. University of Luxembourg.

Chanson, M., Bogner, A., Bilgeri, D., Fleisch, E., & Wortmann, F. (2019). Blockchain for the IoT: Privacy-preserving protection of sensor data. *Journal of the Association for Information Systems, 20*(9).

Dehalwar, V., Kolhe, M. L., Deoli, S., & Jhariya, M. K. (2022). Blockchain-based trust management and authentication of devices in smart grid. *Cleaner Engineering and Technology, 8*.

Dhanda, S. S., Singh, B., & Jindal, P. (2020). Lightweight cryptography: A solution to secure IoT. *Wireless Personal Communications, 112*, 1947–1980.

Elkanishy, A., Furth, P. M., Rivera, D. T., & Badawy, A. A. (2021). Low-overhead hardware supervision for securing an IoT Bluetooth-enabled device: Monitoring radio frequency and supply voltage. *Journal on Emerging Technologies in Computing Systems, 18*(1), Article 6, 28 pages.

Elmaghraby, A. S., & Losavio, M. M. (2014). Cyber security challenges in smart cities: Safety, security and privacy. *Journal of Advanced Research, 5*, 491–497.

Federal Trade Commission. (2015). *Internet of things: Privacy & security in a connected world*. Federal Trade Commission.

Fenga, W., Qin, Y., Zhao, S., & Fenga, D. (2018). AAoT: Lightweight attestation and authentication of low-resource things in IoT and CPS. *Computer Networks, 134*, 167–182.

Gong, L., Alghazzawi, D. M., & Cheng, L. (2021). BCoT Sentry: A blockchain-based identity authentication framework for IoT devices. *Information, 12*, 203.

Huang, Z., & Wang, Q. (2020). A PUF-based unified identity verification framework for secure IoT hardware via device authentication. *World Wide Web, 23*, 1057–1088.

Karakoç, F., Demirci, H., & Harmancı, A. E. (2015). AKF: A key alternating Feistel scheme for lightweight cipher designs. *Information Processing Letters, 115*, 359–367.

Khan, Z., Pervez, Z., & Ghafoor, A. (2014). Towards cloud-based smart cities data security and privacy management. In *Proceedings of the 2014 IEEE/ACM 7th International Conference on Utility and Cloud Computing*. Washington, DC: IEEE.

Kolokotronis, N., Limniotis, K., Shiaeles, S., & Griffiths, R. (2019). Secured by blockchain: Safeguarding Internet of Things devices. *IEEE Consumer Electronics Magazine*.

Kozlov, D., Veijalainen, J., & Ali, Y. (2012). Security and privacy threats in IoT architectures. In *Proceedings of the 7th International Conference on Body Area Networks*. Oslo, Norway.

Kshetri, N. (2017). Blockchain's roles in strengthening cybersecurity and protecting privacy. *Telecommunications Policy, 41*.

Lauria, F. (2017). How to footprint, report and remotely secure compromised IoT devices. *Network Security*.

Li, L., Liu, B., & Wang, H. (2016). QTL: A new ultra-lightweight block cipher. *Microprocessors and Microsystems, 45*, 45–55.

Longo, S., & Cheng, B. (2015). Privacy-preserving crowd estimation for safer cities. In *Proceedings of the 2015 ACM International Joint Conference on Pervasive and Ubiquitous Computing and the 2015 ACM International Symposium on Wearable Computers*. Osaka, Japan.

Mathur, A., Newe, T., Elgenaidi, W., Rao, M., Dooly, G., & Toal, D. (2017). A secure end-to-end IoT solution. *Sensors and Actuators A, 263*, 91–299.
12

McKay, K. A., Bassham, L., Turan, M. S., & Mouha, N. (2017). *NIST IR 8114: Report on lightweight cryptography*. National Institute of Standards and Technology.

Mustafa, G., Ashraf, R., Mirza, M. A., Jamil, A., & Muhammad. (2018). A review of data security and cryptographic techniques in IoT-based devices. In *Proceedings of the International Conference on Future Networks and Distributed Systems (ICFNDS'18)*. Amman, Jordan.

Ragab, A. A. M., Madani, A., Wahdan, A. M., & Selim, G. M. I. (2020). Design, analysis, and implementation of a new lightweight block cipher for protecting IoT smart devices. *Journal of Ambient Intelligence and Humanized Computing*.

Rao, V. V., & Prema, K. V. (2020). DEC-LADE: Dual elliptic curve-based lightweight authentication and data encryption scheme for resource constrained smart devices. *Wireless Sensor Systems*.

Rose, K., Eldridge, S., & Chapin, L. (2015). *The Internet of Things: An overview. Understanding the issues and challenges of a more connected world*. Internet Society.

Satamraju, K. P., & Malarkodi, B. (2020). A PUF-based mutual authentication protocol for Internet of Things. Department of ECE, National Institute of Technology, Tiruchirappalli, Tamil Nadu, India.

Schuß, M., Iber, J., Dobaj, J., Kreiner, C., Boano, C. A., & Römer, K. (2018). IoT device security the hard(ware) way. In *Proceedings of the 23rd European Conference on Pattern Languages of Programs (EuroPLoP '18)* (4 pages). Irsee, Germany.

Sen, B. (2020). PUF: A new era in IoT security. *CSI Publications, Centre for Strategic Infocomm Technologies*.

Seo, H., Jeong, I., Lee, J., & Kim, W. (2018). Compact implementations of ARX-based block ciphers on IoT processors. *ACM Transactions on Embedded Computing Systems, 17*(3), Article 60, 16 pages.

Shamsoshoara, A., Korenda, A., Afghah, F., & Zeadally, S. (2020). A survey on physical unclonable function (PUF)-based security solutions for Internet of Things. *Computer Networks, 183*.

Van Zoonen, L. (2016). Privacy concerns in smart cities. *Government Information Quarterly, 33*, 472–480.

Weber, R. H. (2010). Internet of things – New security and privacy challenges. *Computer Law & Security Review, 26*, 23–30.

Weber, R. H. (2015). Internet of things: Privacy issues revisited. *Computer Law & Security Review, 31*, 618–627.

Witti, M., & Konstantas, D. (2018). A secure and privacy-preserving Internet of Things framework for smart city. In *Proceedings of the International Conference on Information Technology (ICIT 2018)*. Hong Kong, China.

Zhang, K., Ni, J., Yang, K., Liang, X., Ren, J., & Shen, J. (2017). Security and privacy in smart city applications: Challenges and solutions. *IEEE Communications Magazine, 55*(1), 122–129.

Zhang, Y. J., Zhang, L., Yang, F., & Gu, S. T. (2021). Digital fingerprint extraction method of IoT devices based on cryptography. In *Proceedings of the 9th International Conference on Information Technology: IoT and Smart City (ICIT 2021)* (4 pages). Guangzhou, China.

Zhou, L., Su, C., Hu, Z., Lee, S., & Seo, H. (2019). Lightweight implementations of NIST P-256 and SM2 ECC on 8-bit resource-constrained embedded device. *ACM Transactions on Embedded Computing Systems, 18*(3), Article 23, 13 pages.

___________________________________________________________________________